\documentstyle[twoside,fleqn,espcrc2,epsf]{article}

\newcommand{\bsg}{${\rm b}\to {\rm s}\gamma$}
\newcommand{\lsim}{\mathrel{\rlap{\lower4pt\hbox{\hskip1pt$\sim$}}
    \raise1pt\hbox{$<$}}}         
\newcommand{\gsim}{\mathrel{\rlap{\lower4pt\hbox{\hskip1pt$\sim$}}
    \raise1pt\hbox{$>$}}}         
\newcommand{\esim}{\mathrel{\rlap{\raise2pt\hbox{$\sim$}}
    \lower1pt\hbox{$-$}}}         

\title{Direct detection of neutralino dark matter and \bsg\ decays
\thanks{Presented by Paolo Gondolo. 
This work was supported in part by European Community 
Twinning and Mobility (Theoretical Astroparticle Network, TAN) grants.}
}

\author{Paolo Gondolo\address{Universit\'e de Paris VII, Physique
Th\'eorique et Hautes Energies, 2
place Jussieu, 75005 Paris, France}\address{University of Oxford,
Department of Physics, 1 Keble Road, Oxford, OX1
3NP, United Kingdom } and Lars Bergstr\"om\address{Stockholm
University, Department of Physics, Box 6730, S-113 85 Stockholm,
Sweden}\address{University of Uppsala, Department of Theoretical
Physics, Box 803, S-751 08 Uppsala, Sweden}}
       
\begin{document}

\begin{abstract}
We analyze the direct detection rate of minimal supersymmetric
neutralino dark matter in germanium, sapphire and sodium iodide
detectors, imposing cosmological and accelerator bounds including
those from \bsg\ decay. In contrast with several other recent analyses
we find models with light charged higgsinos and large stop mixing in
which the counting rate in solid state detectors exceeds 10
events/kg/day.
\end{abstract}

\maketitle

The recent observation of the \bsg\ decay
\cite{CLEO} has stirred interest in the possible bounds obtainable for
supersymmetric models that contribute to this process
\cite{bertolini,bsgpapers,GaristoNg}. Some authors~\cite{borzumati,bsgDM}
analyzed the consequences of this and other accelerator bounds on the
predicted rates in experiments searching for neutralino dark matter in
the galactic halo. They claimed that allowed rates are small, largely
irrelevant for present-day dark matter detectors.

We performed a detailed study of the allowed supersymmetric parameter
space. In contrast with~\cite{borzumati}, we found models in which the
integrated counting rates are not at all small but as high as to have
already been probed (and excluded) by current negative dark matter
searches. Here we present an overview of our approach and results, and
refer the reader to~\cite{longpaper} for further details.

We work in the framework of the minimal $N=1$ supersymmetric extension
of the standard model (MSSM)~\cite{haberkane}.  We include one-loop
radiative corrections in the Higgs sector according to the effective
potential approach. For simplicity, we make a simple ansatz of a
universal (weak-scale) sfermion mass parameter $m_{\tilde{f}}$.  This
ansatz implies the absence of tree-level flavor changing neutral
currents in all sectors of the model.  Our remaining arbitrary
parameters are the soft supersymmetry breaking trilinear couplings
$A_b$ and $A_t$, the ratio of the two Higgs vacuum expectation values
$\tan\beta$, the pseudoscalar mass $m_{A}$, the higgs(ino) mass
parameter $\mu$ and the gaugino mass parameter $M_2$.  We fix the top
quark mass at $m_t=175$ GeV. We note that this definition of the MSSM
models is the same as in \cite{borzumati}.

We adopt a phenomenological approach and allow for general variations
of parameters in the MSSM, still of course consistent with
experimental bounds and giving correct low-energy symmetry breaking.
We keep only models that satisfy accelerator constraints, including
the 95\% C.L. limits on the \bsg\ branching ratio from the CLEO
experiment \cite{CLEO}, $1.0\cdot 10^{-4} <$ BR(\bsg) $< 3.4\cdot
10^{-4}$. We calculate BR(\bsg) according to~\cite{bertolini}. For a
general supersymmetric model, \bsg\ receives contributions from
W-bosons, charged Higgs bosons, charginos, gluinos and
neutralinos. For our no-FCNC ansatz, the latter two are absent. We
also impose the cosmological constraint $\Omega_\chi h^2 < 1$, where
the relic neutralino density $\Omega_\chi h^2$ is calculated as
in~\cite{GondoloGelmini}.

We compute direct detection rates, integrated over deposited energy
with no energy threshold, for pure germanium (${}^{76}$Ge), sapphire
(Al$_2$O$_3$) and sodium iodide (NaI) detectors. For the local
galactic neutralino velocity distribution we assume a truncated
gaussian of velocity dispersion 120 km/s. We fix the relative
Earth-halo speed at 264 km/s (a yearly average). We adopt a local dark
matter density of 0.3 GeV/cm$^3$ whenever $\Omega_\chi >
\Omega_{\rm gal}$, the minimum value for which neutralinos could
make up the totality of galactic dark matter. When $\Omega_\chi <
\Omega_{\rm gal} $, we scale the local density proportionally to $
\Omega_{\chi} / \Omega_{\rm gal} $.  We use a heavy-squark effective
lagrangian~\cite{efflagrange} for the neutralino-nucleon interaction,
and we use simple gaussian nuclear form factors~\cite{nuclearform},
which are quite adequate for our purposes.

\begin{figure}[htb]
\vspace{-4\baselineskip}
\epsfxsize=90mm \leftline{\epsfbox{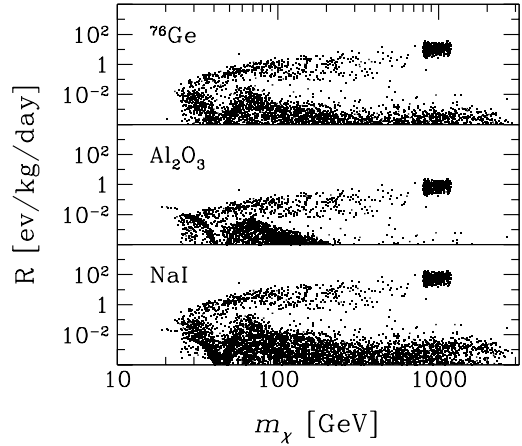}}
\vspace{-5\baselineskip}
\caption{Integrated direct detection rate R in ${}^{76}$Ge,
Al$_2$O$_3$ and NaI detectors versus neutralino mass $m_\chi$.}
\label{fig:3}
\end{figure}

We produce two initial samples of 4000 points each, generating model
parameters randomly within the following bounds: $m_{\tilde{f}} \in
[100,3000]$ GeV, $A_b,A_t \in [-3m_{\tilde{f}},3m_{\tilde{f}}]$, $\tan\beta \in
[1.2,50]$, $m_{A}\le 1000$ GeV, and, for the first sample, $\mu,M_2
\in [-5000,5000]$ GeV, while for the second sample we trade $\mu,M_2$
with the lightest neutralino mass $m_\chi\in[-5000,5000]$ GeV and its
gaugino fraction $Z_g \in [0.00001,0.99999]$.  In these samples, we
find {\it no} points with interesting detection rates, say above 1
event/kg/day in Ge. But we notice that the distributions of points
obtained depends crucially on our choice of sampling.  So we do not
conclude that in our class of models the \bsg\ constraint is so strong
to exclude interesting detection rates. We instead perform two special
scans aiming at large neutralino-proton cross sections. In the first,
we sample in $\mu$--$M_2$ as before but restrict $m_A \in
[0,60]$~GeV. In the second, we sample in $m_\chi$--$Z_g$ with the
restricted $m_A$ range and further demand $m_\chi \in [800,1200]$~GeV
and $Z_g \in [0.01,0.99]$. The results of these special scans for Ge,
Al$_2$O$_3$ and NaI are shown in fig.~\ref{fig:3}, together with (the
top parts of) the initial naive samplings. Remarkably, the high-rate
zones, empty in the initial scans, are now filled with
points. Particularly striking is the concentration around $m_\chi
\esim 1000$~GeV, which obviously comes from the second special
sampling.

Therefore, contrary to~\cite{borzumati}, we find models that do not
violate the experimental and cosmological constraints mentioned above
and in which, thanks to a light neutral Higgs boson, the integrated
counting rate is as large as 10 events/kg/day in Ge and even 100
events/kg/day in NaI.  These models have a relatively light (100--200
GeV) charged Higgs boson, but they are compatible with the \bsg\
constraint because the charged Higgs contribution to the
\bsg\ amplitude is effectively canceled, at large
$\tan\beta$ and large top squark mixing, by the contribution from a
light charged higgsino (cfr.~\cite{GaristoNg}). In terms of the
supersymmetry parameters this occurs when $|M_2| \gg |\mu| \gsim m_W$
and $\mu(A_t+\mu\cot\beta) < 0 $.  We can satisfy these conditions for
both positive and negative values of $\mu$, because we are free to
choose the sign and magnitude of $A_t$. This freedom would be lost in models
imposing additional theoretical constraints, for example in no-scale
models or in models with a flat K\"ahler manifold ($A=0$ or $A=B-m$
respectively at the unification scale).

\begin{figure}[htb]
\vspace{-4\baselineskip}
\epsfxsize=90mm \leftline{\epsfbox{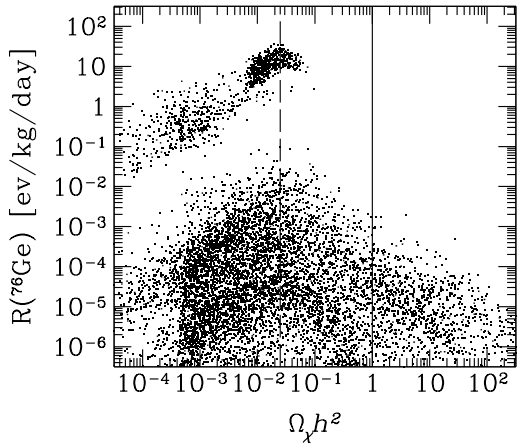}}
\vspace{-5\baselineskip}
\caption{Integrated direct detection rate R off ${}^{76}$Ge
versus neutralino relic density $\Omega_\chi h^2$.}
\label{fig:6}
\end{figure}

In our simple-minded prescription for the local galactic neutralino
density, the value of $ \Omega_{\rm gal} $ is quite uncertain, both
because of uncertainties in the density and extension of galactic
halos and because of the poorly known relation between the
universally-averaged and the local dark matter densities.
Fig.~\ref{fig:6} shows the predicted rates in Ge versus the calculated
neutralino relic density $\Omega_\chi h^2$.  The two initial samples
show at the bottom and the two special samples are the band and cloud
in the upper parts. Models with $\Omega_\chi h^2 > 1$ are plotted only
to illustrate the trend of $R$ versus $\Omega_\chi h^2$. Rates to the
left of the vertical dashed line ($\Omega_\chi = \Omega_{\rm gal}$)
are $\Omega$-suppressed. We have chosen $\Omega_{\rm gal} h^2 =
0.025$, as in~\cite{borzumati}, but the actual value might be even one
order of magnitude larger or smaller. Were it smaller, many models in
which the detection rate is suppressed just because $\Omega_\chi h^2$
is too small would have important detection rates. Notice that in some
of the interesting models, the neutralino relic density is large
enough for them to fill up galactic halos.

We stress that the density of points in the plots depends on the {\it
a priori} distribution of the model parameters, which is entirely at
our choice. One should not be misled in thinking that rate values in
zones where there are more points are more probable than those in
which there are few. This is evident for our special scans, for which
high detection rates with acceptable \bsg\ branching ratios look
``generic.''  No probability should be attached to the plotted point
distributions or to histograms derived from them: the figures can
only illustrate possible neutralino detection rates. 

The conservative approach we propose is to regard the whole range of
outcomes of a calculation as a priori equally probable, irrespective
of the parametrization. This means that really only upper and lower
limits can be given.  Unfortunately, a thorough and fine scanning of
parameter space is computationally very expensive and an alternative
analytical extremization seems prohibitively complicated. We therefore
have to leave the following question open: are there in fact
additional, allowed points in the empty regions of our plots?

\end{document}